
\magnification = 1200
\def\pmb#1{\setbox0=\hbox{$#1$}
\kern-.025em\copy0\kern-\wd0
\kern.05em\copy0\kern-\wd0
\kern-.025em\raise.0433em\box0 }

\vsize=7.5in
\hsize=5.6in
\tolerance 10000

\baselineskip 12pt plus 1pt minus 1pt
\pageno=0
\vskip 8pt
\centerline{{\bf HANDLE OPERATORS OF COSET MODELS}\footnote{*}{This
work is supported in part by funds
provided by the U.S. Department of Energy (D.O.E.) under contract
\#DE-AC02-76ER03069, and by the Division of Applied Mathematics of the U.S.
Department of Energy under contract \#DE-FG02-88ER25066.}}
\vskip 24pt
\centerline{Michael Crescimanno}
\vskip 12pt
\centerline{\it Center for Theoretical Physics}
\centerline{\it Laboratory for Nuclear Science}
\centerline{\it and Department of Physics}
\centerline{\it Massachusetts Institute of Technology}
\centerline{\it Cambridge, Massachusetts\ \ 02139\ \ \ U.S.A.}
\vskip 12pt
\vskip 1.5in
\centerline{Appeared in: {\it Modern Physics Letters\/} {\bf A8} (1993) 1877}
\vfill
\vskip -12pt
\noindent CTP\#2131 \hfil hep-th/9312134 \hfill March 1993
\noindent
\eject \baselineskip 24pt plus 2pt minus 2pt
\centerline{\bf ABSTRACT}
\medskip
Several interesting features of coset models "without fixed points"
are easily understood via Chern-Simons theory. In this paper
we derive explicit formulae for the handle-squashing operator
in these cosets. These operators are fixed, linear
combinations of the irreducible representations of the coset.
As a simple application of these curious formulae, we
compute the traces of all genus-one operators for several common cosets.

\vfill
\eject
\noindent{\bf 1.\quad INTRODUCTION}
\medskip
\nobreak
Although coset models are a relatively old subject$^{1á©á-3}$, they
have recently come under intense scrutiny as settings in which
one may explicitly study properties of string vacua which may
shed some light on, for example, the nature of gravitation
in string theory.$^{4-á©á6}$ Originally coset models were understood
in algebraic terms only and lagrangian descriptions of classes
of coset models, usually in terms of
gauged Wess-Zumino-Witten
(WZW) models, have been studied
extensively in the last few years.$^{7-10}$.

	As well studied as these conformal field theories are,
many structural and even philosophical questions about them
remain unanswered. For example, Witten$^{11}$ recognized the connection
between the Hilbert space of
a Chern-á©áSimons gauge theory in three
dimensions and the conformal blocks of of the $G_k$ theory but,
except for some cursory remarks$^{5,10,12}$, a truly simple and
self contained description of the connection between the general coset
model and some topological three-á©ádimensional theory has yet
to emerge. From such a theory one would learn new
things both about the three-á (such as a
clearer view of the associated link and three-á©ámanifold
invariants) and the two- (for example, a view
of the resolution of "fixed points" under simple current
identification$^{13}$) dimensional theory.

	The aim of this present note is to use the the conceptually
incomplete (although current) view of Cherná©á-Simons gauge theory as it
relates to coset models to answer some structural questions about their
fusion ring. In particular we will derive explicit formula for the
handle-á©ásquashing operator for a class of coset models.
Such formulae arose naturally in studying
the fusion ring of $G_k$ theories.
As the present description of the relation between
Chern-á©áSimons theory
coset theories is really only applicable
to coset theories 'without fixed points' (discussed below), in this
note we will restrict our attention to these theories\footnote{*}{
For experts, we consider "simple" cosets with a single numerator and
denomenator, 'without fixed points' and where the index of embedding is one.}
{}.

	The paper is organized as follows; In section 2 we
review some generalities and methods for describing the space
of conformal blocks of a coset theory.
Section 3 discusses the
handle-squashing operator in coset models.
and section 4 describes one application of these
formulae and a brief conclusion.

\goodbreak
\bigskip
\hangindent=24pt\hangafter = 1
\noindent{\bf 2. \quad Coset Generalities}
\medskip
\nobreak
In an effort to make this note self-á©ácontained we here describe
some basic notions and methodology of cosets. Unfortunately,
this will be a very brief primer: those interested in a more
thorough and systematic introduction are referred to Refs.[5,10]

	Algebraically, coset models may be understood as a variant of the
familiar Kac-©Moody construction of Virasoro representations. The starting
point is to consider the $G_k$ current algebra$^{1-3,14}$,
$$ [ J^a(x),J^b(y)] = f^{abc}J^c(x)\delta(x-y) + k\delta^{ab}\delta'
(x©-y) \ \ , \eqno(2.1)$$
where, as usual, the currents carry
Lie ©algebra indices and
$k$ is some fixed integer. The representation theory for these
algebras is well understood$^{15}$ and the WZW models provide a
rigorous and complete lagrangian description of these models$^{4}$.
It was noticed long ago that one could find a representation
of the Virasoro algebra
on a subspace of the Hilbert space of the original model
that is annihilated
by a subset of the currents.
The motivation for this approach arose
originally from trying to understand
confinement in the strong interactions
from a string theory point-of-view.
At any rate, for the
resulting theory to be unitary it is necessary to require that the
chosen subset of currents close algebraically; that is, they must form a
subalgebra of Eq.(2.1). This subalgebra is itself usually a Kac-Moody algebra
associated with a subgroup $H$ of the original group $G$.
Requiring these chosen currents to vanish on all the states
results in a description of the coset model's Hilbert space.
The resulting theory is denoted by $G/H$. The representation theory
of such coset theories is well understood$^{2,3}$ and
one builds a Virasoro representation from this model
via the Suguwara$^{16}$ construction, in which the stress tensor of the
coset is realized as the difference of the stress tensors of the
$G$ theory and the $H$ theory. These simple cosets have a
lagrangian formulation in terms of gauged WZW models$^{7-9}$.

	In heuristic terms, it is often possible to describe (a
basis for) the vector space of the conformal blocks of the
coset theory $G/H$ in terms of the individual conformal blocks of the $G_k$
and the $H_k$ theory. This may be understood in terms
of simple considerations on the representation spaces
$W_G$ and $W_H$ of the
Lie algebras of $G$ and $H$.
Given a representation $R$ of the
Lie algebra $G$, we may decompose it with respect to the
representations $r$ of the subgroup $H$,
$$R = \sum_{r}^{} b_r^R \ r \ \ .   \eqno(2.2)$$
In decomposing a given representation $R$ into
$H$ representations, the R.H.S. of Eq.(2.2) does not
contain every representation. The representation theory
of $G$ (and thus $G_k$) is graded with respect to the action of $Z_G$,
the center of the group. Let $Z=Z_G\cap H$, the common centers
of $H$ and $G$. Thus for $b_r^R\ne 0$ it is necessary
that $R$ and $r$ have the same $z$ eigenvalue for each $z\in Z$.
That is,
$$ b_r^R \in (W_G \otimes W_H)^Z\ \ .   \eqno(2.3)$$
where by $(W_G \otimes W_H)^Z$ we mean the $Z$-invariant
part of the vector space $W_G \otimes W_H$. We will see
below how this 'selection rule' is applied in
simple coset models.

The representations of Kac-Moody algebra $G_k$
above are labeled by a subset of the representations
of the Lie algebra $G$.
This subset is called the integrable representations
and they
form a vector space that we shall, following
Ref.[10], denote by $V_G$.
One may also decompose Kac-Moody representations
with respect to a subalgebra (i.e. $H_k$) and in simple
cases an equation like Eq.(2.2) is recovered, where now the sum
is over the integrable representations of the algebra $H_k$.
The operation of finding the subspace of the
Hilbert space of the $G_k$ which is annihilated by the
$H_k$ currents is equivalent to setting all the $r$'s
in the R.H.S. of Eq.(2.2) to one. That is (in simple cases atleast)
the $b_r^R$ correspond to the integrable representations or
"current blocks"
of the $G_k/H_k$ model.

Chern-Simons theory
is a three-dimensional gauge theory that provides a complete description
of the space $V_G$ and of the linear operators of interest on
it. More than just a philosophical framework,
Chern-Simons theory
can be used for explicit computation and has indeed
been useful in elucidating the structure of $G_k$ as a
conformal field theory$^{11,17-20}$. However, a
corresponding three-dimensional viewpoint for coset
theories and for more general conformal field theories is
somewhat incomplete. For example, in $G_k$, the three-dimensional
theory is understood in terms of quantizing the moduli
space of flat $g$-connections ($g$ is the Lie algebra of $G$);
for the general conformal model no correspondingly
simple picture has emerged.

For a certain simple class of coset models there is a recipe
for how to proceed$^{5,10}$. Motivated by Eq.(2.2), one identifies
with the coset conformal blocks the subspace of $V_G \otimes \hat{V}_H$
($V_G$ is the space of conformal blocks of the theory $G_k$, and
the $\hat{V}$ means the dual of the vector space $V$)
invariant under the action of the common center $Z=Z_G\cap H$.
This recipe admits a Chern-Simons interpretation: one requires the
Wilson line operators associated to the common center $Z$ to
be trivial (i.e. =1) in the coset model's vector space of current blocks.
We denote the set of Wilson line operators associated with the
center action as $Z'= Hom(H_1(\Sigma), Z)$.
It is to be stressed that this is still
essentially a recipe and indeed works completely and unambiguously
only when the orbits of the $Z'$ action in $V_G \otimes \hat{V}_H$
all have the same length. For a generic coset this is not
the case, and when all the orbits are not of the same length
the construction is referred to as a coset
'with fixed points'$^{13,21-26}$.
Nonetheless, cosets in
which the above mentioned orbits all
have the same length (called cosets 'without fixed points')
do play a prominent part
in conformal field theory. For example,
the cosets $SU(2)_k \over U(1)_k$
and $SU(2)_k\times SU(2)_l \over SU(2)_{k+l}$ ($l$ odd)
are cosets
'without fixed points'.
\goodbreak
\bigskip
\hangindent=24pt\hangafter =1
\noindent{\bf 3.\quad Handle-Squashing in Cosets}
\medskip
\nobreak
For coset models 'without fixed points' (as discussed in the
previous section) the modular transformation $S\ :\ \tau \rightarrow
{{-1}\over{\tau}}$ on the torus is realized as a unitary operator on the
space of blocks
$$ S_{G/H} = S_G \otimes {S_H}^{\dagger}\ |_{(V_G \otimes \hat{V}_H)^{Z'}} \ \
{}.
\eqno(3.1) $$
This $S$ matrix is obviously unitary on $(V_G \otimes \hat{V}_H)^{Z'}$
and, for cosets "without fixed points"
it leads (via the Verlinde formula) to integer
fusion coefficients. Indeed, the fusion algebra of such cosets is simply
given via tensoring operators in the original theories. Using
the canonical map in the conformal field theory
that
relates states
to operators (through ${\cal O}_\Lambda \psi_0 = \psi_\Lambda$ )
$$ V_G \rightarrow End(V_G,V_G)\ \ ,  \eqno(3.2)$$
we find an identification of the operators on $V_{G/H}$
$$ End(V_{G/H},V_{G/H}) = Image\big\{ (V_G\otimes\hat{V}_H)^{Z'} \rightarrow
End(V_G,V_G)\otimes End({\hat V}_H,{\hat V}_H) \big\}  \ \ .
\eqno(3.3)$$

Recall that
in the case of simple Kac-Moody conformal field theory $G_k$
there is an explicit formulae$^{19}$ for the inverse of the
$K$-matrix$^{27,28}$ in terms of a fixed linear combination
(i.e. the coefficients were independent of the level)
of the integrable representations.
In Ref.[19] formulae for the handle-squashing operator
, $K^{-1}$, arose by
writing the inner product in the space $V_G$ in terms
of the associated Gaussian model\footnote{$\dagger$}{
	The Gaussian model is isomorphic to the
	ring of theta functions from which one
	forms the characters of the theory.}
{}.
Philosophically, this is the strict conformal field theory
analogy of the functional norm for the polynomial representation
of the fusion algebra introduced in Ref.[29] (for additional background
see Refs.[19,30-34].)
The $K^{-1}$ matrix defines a
norm on the space of operators of the conformal field
theory,
$$ \delta_{ij}=Tr({\cal{O}}_{\overline{i}}{\cal{O}}_{j} K^{-1})
\ ,\eqno(3.4)$$
where $Tr$ denotes trace over space of conformal blocks at genus one.
For a coset of the type mentioned
above ('without fixed points') we now show that there is also a
simple formula for the handle-squashing operator that descends
from the constituent $G_k$ theory.

The space of
conformal blocks (in genus one)
for the coset is isomorphic to $(V_G \otimes \hat{V}_H)^{Z'}$
where $Z'=Hom(H_1(T^2),Z)\approx Z\times Z$ is the abelian group of the
operators in the fusion algebra that are associated with the
action of the common center $Z=Z_G \cap H$. As described above,
these operators are always 'simple' currents
$^{13,22-25}$
and, by construction, generate an automorphism of
the fusion algebra. Let $|Z|$ be the order of the group $Z$
and let $z_l\ \in\  Z'$ be
the operator
corresponding to transport along cycle $l$(=1,2, labels of
the homology basis)
of the torus
in some representation associated with the
center.

	Thus, if $Z'$ acts without fixed points the $S$-matrix
(associated to modular transformations of the torus) is given as in
Eq.(3.1). This is described in Refs.[21-26] where
it is shown that
projecting the matrix $S_G\otimes S_H^\dagger$
onto the $Z'$ invariant states
results in a simple normalization factor $|Z|$ of the
rows of the $S_G\otimes S_H^\dagger$ matrix. Now, since the eigenvalues of
the $K$ matrix of Verlinde$^{27}$
(see also Bott$^{28}$)
may be written as $|S_{0i}|^{-2}$, the  handle-squashing
operator $K^{-1}$ of the coset is,
$$ {K_{G/H}}^{-1} = |Z|^2\ \big( {K_G}^{-1}\otimes{K_H}^{-1}\big)
\big|_{G/H}\ \ .
\eqno(3.5)$$

	Another derivation of this simple
result which doesn't make explicit use of Eq.(3.1) but proceeds
from analogy with the original
derivation of handle-squashing operators in
the $G_k$ theory is given in appendix A. That is,
the derivation relates the inner product on the space of operators of the
coset theory with the inner product on the space of operators
of the Gaussian model.

	Actually, it is possible to make a stronger statement about
the form of the handle-squashing operator.
Appendix A contains a proof that for $G_k$,
the handle-squashing operator ${K_G}^{-1}$
commutes with all the operators $z\in Z'$ of the center representations.
Now, as shown in Ref.[19], ${K_G}^{-1}$ is proportional to
a particular linear combination (independent of the level)
of the integral representations of
the $G_k$ theory. Since it commutes with every $z\in Z'$
then under the projection to the coset $\big |_{G/H}$
it follows that the handle-squashing
${K_{G/H}}^{-1}$ of the coset is itself
proportional to a particular linear combination (independent of the level)
of the representations
of the coset model.

As a particularly simple example of the formula Eq.(3.5)
consider the coset $U(1)_k\times U(1)_l/U(1)_{k+l}$. For
$U(1)_k$ the $K^{-1} = {{1}\over{2k}} {\cal{O}}_1$, where ${\cal{O}}_1$
is the identity representation ($i.e$ for $U(1)_k$ there are $2k$ blocks and
in this convention, we assign the label '$1$' to the
trivial representation, the unit matrix)
Now, following
Ref.[5], we understand the center of $U(1)_k\times U(1)_l/U(1)_{k+l}$
to be $Z = {\bf Z}_{2(k,l)}$ where ${\bf Z}$ is the integers and
$(k,l)$ is the greatest common divisor of $k$ and $l$. Thus
using Eq.(3.5) we find that for this coset,
$$ {\rm for}\  {U(1)_k\times U(1)_l}/U(1)_{k+l} \qquad
K^{-1} = {{(k,l)^2}\over{2kl(k+l)}}{{\cal{O}}_{1,1,1}} \ .  \eqno(3.6)$$
where ${\cal{O}}_{1,1,1}$ is simply the unit matrix (the
trivial representation) on the coset's space of blocks. Eq.(3.6)
is precisely what one would expect from the equivalence,
$${{U(1)_k\times U(1)_l}\over{U(1)_{k+l}}} = U(1)_{{kl(k+l)}\over{(k.l)^2}}
\ \ .  \eqno(3.7)$$

For a less trivial example, consider the coset $SU(2)/U(1)$ at level k. The
``vacuum'' state of this coset model in terms of
$V_{SU(2)}\otimes\hat{V}_{U(1)}$ states is
$$ |0>_{SU(2)/U(1)}={
{({\cal{O}}_1\otimes{\cal{O}}_1 +
{\cal{O}}_{k+1}\otimes{\cal{O}}_{k+1})}
\over
{\sqrt{2}}}\ \
|0>\otimes<0|\ ,
\ \ \eqno(3.8)$$
where the subscripts on the $\cal(O)$ signify the dimensions
of the representation and $|0>\otimes<0|$ is the tensor product
of the vacuum of the $SU(2)_k$ and $U(1)_k$ theories respectively.
Using the fact that the associated
$K^{-1}$ matrices
are,
$$ K^{-1}_{SU(2)}={1\over{2(k+2)}}(3{\cal{O}}_1-{\cal{O}}_3) \eqno(3.9)$$
and
$$ K^{-1}_{U(1)}={1\over{2k}}{\cal{O}}_1 \eqno(3.10)$$
with the common center $Z={\bf Z}_2$ we learn,
$$ K^{-1}_{SU(2)/U(1)} ={1\over{k(k+2)}}(3{\cal{O}}_1\otimes{\cal{O}}_1-
{\cal{O}}_3\otimes{\cal{O}}_1)\ .\    \eqno(3.11)$$
Separately, each term in the above expression
is an ${\cal{O}}\in End(V_{G/H},V_{G/H})$ as may readily be
seen by applying each of them to the vacuum state
$|0>$ of Eq.(3.8).
We label the operators of the coset
as allowed (according to the selection rule discussed above, that is,
invariance with respect to the action of $Z'$)
products of operators of the constituent theories
and recall that this labelling (and fusion) is
defined modulo
the action of the center $Z'$.
Thus, in general $K_{G/H}^{-1}$ is a proportional to a fixed
(independent of the level $k$) linear combination of
operators of the coset, as expected
from the fact that $K^{-1}$ commutes with all $z\in Z'$.

In the next section we
describe the explicit formulae for the handle-squashing operator
in several other simple theories and show how it may be
used to compute the traces of operators in the theory.

\goodbreak
\bigskip
\hangindent=24pt \hangafter= 1
\noindent{\bf 4.\quad Trace Formulae and Conclusion}
\medskip
\nobreak

One intriguing application of the explicit formulae
for the handle-squashing operators is its use in
finding the traces of operators.
Using
Eq.(3.4) above, we find $Tr(K^{-1})=1$ and
$Tr(K^{-1}{\cal{O}}_j)=0$ for
all operators ${\cal{O}}_j$ not the identity. We now
turn these relations into formulae for the traces
of the individual operators of the theory.

The
trace $Tr({\cal{O}}_j)$, where the trace is taken over
the space of conformal blocks of genus one, is
simply the number of one-point blocks on the torus
with the label $j$.
Explicit formulae for the traces are useful for,
among other things,
computing the handle operator ($K$-matrix)
$$K = \Sigma_{l}^{}\ Tr({\cal{O}}_l){\cal{O}}_l \ . \eqno(4.1)$$
Thus, the traces are always integer
(in general depending on the level, etc.), and
usually are found combinatorially directly from the
fusion rules. The point we wish to make here is that
using the explicit form of the handle-squashing operator
it is possible to compute these operator traces without
resorting to combinatorics.

This is particularly simple for $SU(2)_k$. Note
first that $Tr({\cal{O}}_j)=0$ for all $j$ even
integer (half-odd integer spins operator).
This follows from the fact that there are $z\in Z'$
for which $z{\cal{O}}_j z^{-1} = (-1)^{j+1}{\cal{O}}_j$
and
thus implies that $Tr({\cal{O}}_j)=0$ for
all even $j$. In general, for any $G_k$, those representations
which are charged under the center action (for $SU(N)$, this
means all those operators with non-trivial $N$-ality) have
zero trace.

The trace of the integer spin operators in $SU(2)_k$
are easy to find. Using the explicit form
for the $K^{-1}$ (Eq.(3.7) above) for $SU(2)_k$ and
the condition $Tr(K^{-1})=1$ one finds $Tr({\cal{O}}_3)=k-1$
and by using $Tr(K^{-1}{\cal{O}}_j) =0$, $j\ne 1$ and
the fusion algebra, one
easily finds,
$$ {\rm for}\  SU(2)_k \qquad Tr({\cal{O}}_j)=k+2-j \qquad j\  {\rm odd}
\ . \eqno(4.2)$$

A similar analysis of the coset $SU(2)_k/U(1)_k$ yields,
$$ {\rm for}\  SU(2)_k/U(1)_k \qquad
Tr(m,1)={{k(k+2-m)}\over{2}}
\qquad m\ {\rm odd}\ . \eqno(4.3)$$
with the notation $(m,l) = {\cal{O}}_m\otimes{\cal{O}}_l$, where
the first factor corresponds to the $SU(2)_k$ label and the second to the
$U(1)_k$ label. Due to the center symmetry the labels $(m,l)$
run from $m=1,2,3,...,k+1$ and $l=1,2,3,...,k$
with $m$ and $l$ either both
even or both odd.
All other operators have zero trace (the trace is, of course, taken
over the blocks of the coset model on the torus.)

The models $SU(2)_k\times SU(2)_l/SU(2)_{k+l}$
are an interesting class of cosets related
to the supersymmetric and non-supersymmetric minimal models.
In order for this to be a coset without fixed points it is necessary
that one of either $k$ or $l$ be an odd integer thus,
without loss of generality, in what follows we assume that
$l$ is odd.
To fix notation, we will label each operator
(and thus each state) by a triple of integers
$(n,m,p)$ each number being the dimension of the
representation of the corresponding $SU(2)_k$, $SU(2)_l$
and $SU(2)_{k+l}$ factor respectively.
Obviously, the action of the center furthermore implies that
the labeling is either $(odd,odd,odd)$ and $(odd,even,even)$
for $k$ odd, {\bf or} $(odd,odd,odd)$ and $(even,even,odd)$ for
$k$ even.
It is relatively straightforward using the ideas
discussed above and the explicit form of the
handle operator in these models to compute
the traces of all the operators.
A trivial selection rule (the traceslessness of the
spinor representations) implies that only the trace of the
operators associated with the $(odd,odd,odd)$ sector are
nonzero. Finally, using the ideas discussed in the body of the
paper, we compute these traces without resorting to combinatorics
to find,
$${\rm for}\ {{SU(2)_k\times SU(2)_l}\over{SU(2)_{k+l}}}
\quad Tr(m,n,p)=(k+2-m)(l+2-n)(k+l+2-p)/4\ \qquad m,n,p\ odd. 	\eqno(4.4)$$

It is relatively straightforward, however tedious, to compute the
traces of operators for many other
models at any level with these explicit formulae
of the handle operators.

In conclusion, in this note we have shown that many
common coset models have handle-squashing operators
that are simply related to the constituent theory's.
It would be clearly of much interest to extend this
derivation (and the methodology of the quantization
of moduli space)
to cosets 'with fixed points' and other, more general, models
although first a more thorough understanding
the 'fixed point resolutions' of
ref.[13] in the context of Chern-Simons theory
is necessary.

\goodbreak
\bigskip
\hangindent=24pt \hangafter=1
\noindent{\bf 6. \quad Acknowledgements}
\medskip
\nobreak

The authors gratefully acknowledge conversations with S. Axelrod,
K.Bardakci, D.Freed, S. Elitzur, S.A. Hotes and I.M. Singer.

\vskip 140pt
\noindent{\bf Figure 1} {A diagramatic view of why $[K,z]=0$.}
\vfill
\eject

\centerline{\bf REFERENCES}
\medskip
\item{1.}K. Bardakci and M. B. Halpern, {\it Phys. Rev.\/} {\bf D3} (1971)
2493.
\medskip
\item{2.}P. Goddard, A. Kent and D. Olive, {\it Phys. Lett.\/} {\bf B152}
(1985)
\medskip
\item{3.}K. Bardakci, E. Rabinovici and B. S\"aring, {\it Nucl. Phys.\/}
{\bf B299} (1988), 151.
\medskip
\item{4.}D. Gepner and E. Witten, {\it Nucl. Phys.\/} {\bf B278} (1986), 493.
\medskip
\item{5.}G. Moore and N. Seiberg, {\it Phys. Lett.\/} {\bf B220} (1989), 422.
\medskip
\item{6.}E. Witten, {\it Phys. Rev\/} {\bf D44} (1991), 314.
\medskip
\item{7.}D. Karabali, Q.-H. Park, H. J. Schnitzer and Z. Yang,
{\it Phys. Lett.\/} {\bf 216B} (1989) 307;\ H. J. Schnitzer, {\it Nucl.
Phys.\/} {\bf B324} (1989) 412;\ D. Karabali and H. J. Schnitzer,
{\it Nucl. Phys.\/} {\bf B329} (1990) 649.
\medskip
\item{8.}K. Bardakci, M. Crescimanno and E. Rabinovici, {\it Nucl. Phys.\/}
{\bf B344} (1990), 344.
\medskip
\item{9.}K. Gawedzki and A. Kupiainen,
{{\it Phys. Lett.\/} {\bf B215} (1988), 119};
{{\it Nucl. Phys.\/} {\bf B320} (1989) 625.}
\medskip
\item{10.}E. Witten, {\it Commun. Math. Phys.\/} {\bf 144} (1992), 189.
\medskip
\item{11.}E. Witten, {\it Commun. Math. Phys.\/} {\bf 121} (1989), 351.
\medskip
\item{12.}J.M. Isidro, J.M.F. Lambastida and A.V.Ramallo, {\it Phys. Lett.\/}
{\bf B282} (1992), 63.
\medskip
\item{13.}A.N. Schellekens and S. Yankielowicz, {\it Int. J. Mod. Phys.\/}
{\bf A5} (1990), 2903.
\medskip
\item{14.}E. Witten, {\it Commun. Math. Phys.} {\bf 92} (1984), 455.
\medskip
\item{15.}V.G. Kac and M. Wakimoto, {\it Adv. in Math.\/} {\bf 70} (1988), 156.
\medskip
\item{16.}H. Suguwara, {\it Phys. Rev.\/} {\bf 170} (1968), 1659.
\medskip
\item{17.}S. Elitzur, G. Moore, A. Schwimmer and N. Seiberg, {\it Nucl.
Phys.\/}
{\bf B326} (1989), 108.
\medskip
\item{18.}M. Crescimanno and S.A. Hotes, {\it Nucl. Phys.\/} {\bf B372}
(1992), 683.
\medskip
\item{19.}M. Crescimanno, {\it Nucl. Phys.\/} {\bf B393} (1993), 361.
\medskip
\item{20.}C. Imbimbo, "$Sl(2,{\bf R})$ Chern-Simons Theories with Rational
Charges and 2-Dimensional Conformal Field Theories," Genoa Preprint
GEF-TH 5/1992, Hepth/9208016.
\medskip
\item{21.}D. Gepner, {\it Phys. Lett.\/} {\bf B222} (1989), 207.
\medskip
\item{22.}K. Intriligator, {\it Nucl. Phys.\/} {\bf B332} (1990), 541.
\medskip
\item{23.}R. Brustein, S. Yankielowicz and J.B. Zuber,
{\it Nucl. Phys.\/} {\bf B313}
(1989), 321.
\medskip
\item{24.}B. Blok and S. Yankielowicz, {\it Nucl. Phys.\/} {\bf B315} (1989),
25.
\medskip
\item{25.}A.N. Schellekens and S. Yankielowicz, {{\it Nucl. Phys.\/} {\bf B327}
(1989), 673}; {{\bf B334} (1990), 67}.
\medskip
\item{26.}C. Ahn and M.A. Walton, {\it Phys. Rev.\/} {\bf D41} (1990), 2558.
\medskip
\item{27.}E. Verlinde and H. Verlinde, "Conformal Field Theory and
Geometric Quantization," published in
{\bf Trieste Superstrings} (1989), 422.
\medskip
\item{28.}R. Bott, {\it Surveys in Diff. Geom.\/} {\bf 1} (1991), 1.
\medskip
\item{29.}D. Gepner, {\it Commun. Math. Phys.\/} {\bf 141} (1991), 381.
\medskip
\item{30.}K. Intriligator, {\it Mod. Phys. Lett.\/} {\bf A6} (1991), 3543.
\medskip
\item{31.}M. Bourdeau, E. Mlawer, H. Riggs, and H. Schnitzer,
{\it Mod. Phys. Lett.\/} {\bf A7} (1992), 689.
\medskip
\item{32.}D. Nemeschansky and N. P. Warner, {\it Nucl. Phys. \/}{\bf B380}
(1992), 241.
\medskip
\item{33.}D. Gepner and A. Schwimmer, "Symplectic Fusion Rings and
their Metric," Weizmann Preprint WIS-92/34,  Hepth/9204020.
\medskip
\item{34.}P. Di Francesco and J.-B. Zuber, "Fusion Potentials I," Saclay
preprint SPhT 92/138, Hepth/9211138.

\vfill
\eject
\noindent{\bf APPENDIX A. Derivation of
handle-squashing operator for cosets}
\bigskip

	Here we give another derivation of Eq.(3.5) by
studying the inner product on the space of operators
of the theory. This inner product is induced from the
canonical inner product on the space of states $V_G$ by the
1-to-1 correspondence with the operators in the theory.

	We proceed as in the $G_k$ case by first relating the
vacuum state of the coset model to the vacuum of the Gaussian
model
$$ \psi_0 = \tilde{\Gamma} |0>  \qquad \tilde{\Gamma}=\sqrt{N} P_0 \Gamma\ \ ,
\eqno(A.1)$$
where $|0>=|0>_G\oplus\ _H<0|$ is the vacuum state of the Gaussian model
associated to $G_k\otimes H_k$ and
$\Gamma = \Gamma_G\otimes{\Gamma}^\dagger_H$ where $\Gamma_G$ is the
operator that relates the vacuum of the Gaussian model
to the vacuum of $G_k$ (we use the notation and convention
of Ref.[18,19] throughout.) $P_0$ is a projection operator
that annihilates any state that is not invariant under the action
of every $z_2 \in Z'\sim Z\times Z$
where by $z_2$ we mean the center action written in terms
of raising operators of the quantization (as is $\Gamma$ in
Eq.(A.1).)\footnote{*}{By the subscript $j$ on $z_j$ we
mean a label of the homology basis of the torus.}
For example, if $Z$ is ${\bf Z}_N$
then $P_0 = {1\over N}\sum_{j=0}^{N-1} {z_2}^j$
where ${z_2}$ is a generator of ${\bf Z}_N$.
In general $Z$  is simply abelian and finite
and not simply ${\bf Z}_N$ for some $N$, but
it is always simple to construct $P_0$ and
the discussion here won't rely on any particular
form of $P_0$ or $Z$.

Let ${\cal O}_{l}$
be the operator in the coset theory associated to the
coset state labeled by the (multi-index) $l$, that is,
$\psi_l = {\cal O}_{l} \psi_0$. Then we may write
the inner product on the coset states as
$$ (\psi_l,\psi_m) = \delta_{lm} ={1\over{\lambda_G \lambda_H}}
Tr_{Gauss}({\cal O}_{\bar l}{\cal O}_m
{\tilde{\Gamma}}^\dagger {\tilde{\Gamma}}) \ \ ,
\eqno(A.2)$$
where $Tr_{Gauss}$ means that the trace is to be taken over the
Gaussian model's state space (of the $G_k\otimes H_k$ theory)
and
$\lambda_G \equiv \big|{\Lambda_w \over{(k+c)\Lambda_r}}\big|$ where
$\Lambda_w$ and $\Lambda_r$ are, respectively,
the weight (actually the co-root) lattice
and root lattice of the Lie-algebra of $G$.
Let $|W_G|$ be the order of the Weyl group of $G$.
Using the definition of ${\tilde{\Gamma}}$ in Eq.(A.1) we have
$$ \delta_{lm}= {{|Z|\ |W_G||W_H|}\over {\lambda_G \lambda_H}}
Tr_{(V_G\otimes \hat{V}_H)^{P_0}}({\cal O}_{\bar l} {\cal O}_m
{\Gamma}^\dagger\Gamma)\ \ ,
\eqno(A.3)$$
where $(V_G \otimes \hat{V}_H)^{P_0}$ stands for the subspace of
$V_G\otimes\hat{V}_H$ stabilized by $P_0$.

	In order to write Eq.(A.3) as a trace over just the
states in the coset we must understand the action of the remaining
elements of the group $Z'$, that is, the $z_1$ on $V_G\otimes\hat{V}_H$.
These are
the operators that assign a phase to each of the states
of $V_G\otimes\hat{V}_H$. Any ${\cal O}_{l,2}$
(not necessarily associated with a state in the coset)
is homogeneous
with respect to the action of the $z_1\in Z'$ because it is
associated with a irreducible representation and
so must commute {\it up to a phase} (the center is abelian, and these are
the one-dimensional representations)
with each $z_1 \in Z'$.
However each particular ${\cal O}_{l,2}$  associated with a state in the coset
commutes with each $z_1\in Z'$ (without phases) by construction.

	Finally if each $z_1\in Z'$ commuted with the operator
${\Gamma}^\dagger\Gamma$ then the trace in Eq.(A.3) would break up into
sums of traces over individual eigenspaces labeled by
different $z_1$ eigenvalues. We now show that this is indeed the case.
Because
${\Gamma}^\dagger\Gamma$ is ${\Gamma_G}^\dagger\Gamma_G\otimes{\Gamma_H}
{\Gamma_H}^\dagger$
it will be enough to show that each $z_1$ commutes with
${\Gamma_G}^\dagger\Gamma_G$. Now note that ${\Gamma_G}^\dagger\Gamma_G$ is
proportional to the $K^{-1}$ of the $G_k$ theory. Each $z_1$
is a grading (indeed a one-dimensional representation of an
automorphism) of the fusion algebra. As such it is easy to see
diagramatically (see figure 1)
that $z_1$ and $K$ must commute. Thus so do $z_1$ and $K^{-1}$.
For the reader who is unsatisfied with this diagramatic proof,
appendix B contains an explicit Lie-algebraic proof of the
commutation of $z_1$ and $K^{-1}$. At any rate, it is clear
from these arguments that
$$ \delta_{lm} = {{|Z|\ |W_G||W_H|}\over{\lambda_G \lambda_H}}
\sum_s^{} Tr_{{\cal H}_s}({\cal O}_{\bar l} {\cal O}_m {\Gamma}^\dagger
\Gamma)\ \ ,
\eqno(A.4)$$
where ${\cal H}_s$ are the eigenspaces of the
various $z_1\in Z'$ ($s$ is a multi-index distinguishing the
the various eigenvalues of the $z_1$'s) in
$(V_G\otimes\hat{V}_H)^{P_0}$. Note that since the $z_1$'s
are 1-dimensional representations of an automorphism of the
algebra their various eigenspace are distinct and orthogonal.
Furthermore, there are exactly $|Z|$ of these spaces.
The coset $(V_G\otimes\hat{V}_H)^{Z'}$ corresponds to
the ${\cal H}_1$, i.e. the subspace on which
all the eigenvalues of the $z_1$'s
are 1.

Let us now assume that the automorphism group $Z'$ acts 'without fixed points'
as described earlier. Then  each
orbit under the $z_2\in Z'$ in of length $|Z|$.
Since  the operators inside the trace of Eq.(A.3) commute with all  the $z\in
Z'$ and each state occurs once on the orbit under the
automorphism group we
learn that each term in the above sum is the same. Thus
$$ \delta_{lm} = {{|Z|^2
|W_G||W_H|}\over {\Lambda_G \Lambda_H}} Tr_{G/H}({\cal O}_{\bar l}{\cal O}_m
{\Gamma}^\dagger\Gamma)\ \ , \eqno(A.5)$$
and so
$$ {K_{G/H}}^{-1} = |Z|^2\ \big(
{K_G}^{-1}\otimes{K_H}^{-1}\big)\big|_{G/H}\ \ , \eqno(A.6)$$
is the handle-squashing operator of the coset
in  terms of the handle-squashing operators of
the constituent  $G_k$ theories. Note that this derivation made critical use of
the assumption that $G/H$ is a coset 'without fixed points'. This result is
consistent with Eq.(3.1) as the operators $K^{-1}$ have a simple interpretation
in terms  of $|S_{0l}|^2$.
\vfill
\eject
\noindent{{\bf APPENDIX B. Lie Algebraic proof
of $[\Gamma^\dagger\Gamma,z_1]=0$}}
\bigskip
It is relatively straightforward, using ideas from the quantization of
Chern-Simons theory to show directly
that the center action always commutes
with the $K^{-1}$ matrix.
In this appendix we first show that there will be a unique
set of fields in the $G_k$ that will carry the full center symmetry
for any $G$ and level $k$. Having demonstrated existence, we then
directly show that the operators generating the center
commute with $K^{-1}$. This approach has the advantage that
one never explicitly uses the fact that this
symmetry generates an automorphism of the fusion rules.

To show that there will always be a unique set of fields that
generate the full center symmetry of $G$ in $G_k$, we begin by
recalling that, from the point of view of quantizing the moduli space of
flat $g$-connections over $T^2$, the states of $G_k$ are
odd under the generators of the Weyl group whereas the operators
are even under Weyl. It is always possible to find a monomial
in the raising operators of the Gaussian model that is Weyl invariant.
Let $s_l$ be the components of a $rank(G)$ vector that
correspond to such a monomial. That is, consider the monomial
$$ z_2 = \Pi_{j} B_{j}^{s_j}\ .\eqno(B.1)$$
Requiring this monomial to be Weyl invariant gives a condition on the
$s_l$,
$$ exp\big({{2\pi i}\over{k+c}} A_{jm}{C^{-1}}_{ml}s_l\big)=1\qquad \forall j
\ \ . \eqno(B.2)$$
where $A$ is the Cartan matrix and $C$ is the matrix associated
to the Lie-algebraic part of the symplectic form of the quantization. As
described in Refs.[18,19], for simply laced groups, $C$ is simply the
Cartan matrix (more generally, $C$ is the matrix of inner products
of the root vectors and so is always symmetric, even for non-simply
laced $G$.) For clarity of this exposition we discuss the simply laced
case, and note here that it is straightforward to extend this discussion to
include the non-simply laced case. Thus the condition that $z_2$ be
Weyl invariant is simply that
$$ s_l = 0 \ mod(k+c)\ \ . \eqno(B.3)$$
We call the primitive $z_2$ those for which $s_l$'s are zero except
for a single component equal to $k+c$. Some of these primitive $z_2$ will
not be $1$ in the Gaussian model. It is easy to see that there will
be precisely one non-trivial primitive monomial for each element in the
center of $G$.

Now, since these primitive $z_2$ are Weyl invariant, they must take
Weyl-odd states to other Weyl-odd states (by Weyl-odd, we mean
those states odd under each of the generating reflections of the
Weyl group.) Now, the Weyl-odd states are precisely the states
that represent the current blocks of the $G_k$ theory and since
each of these Weyl-odd states are composed of distinct linear
combination of Gaussian states, the monomial operator $z_2$
can take each Weyl-odd state to precisely one other Weyl-odd state,
up to an overall phase. Furthermore, such a $z_2$ may be
interpreted as the fusion operator associated to the state
proportional to $z_2 \psi_0$, where $\psi_0$ is the vacuum state
of the Weyl-odd states. Thus, there are, for every $k$, states
whose fusion realizes the center action are simple
currents in the sense of Refs.[13,21-25].

Having shown that there is always a realization of the center
of the group in terms of a particular set of operators of the
theory, we now show that these operators must commute
with $K^{-1}$, the handle-squashing operator. As advertised
we show this in complete generality with a simple computation.

We proceed as follows; We first show that $[\Gamma,z_1]$ has a
simple form and then use this to compute $[\Gamma^\dagger\Gamma,z_1]$.
{}From the point of view of the quantization of Chern-Simons theory
we describe this entirely in Lie algebraic terms. Recall that $\Gamma$,
is the raising operator in the Gaussian model that applied to
$|0>$ of the Gaussian model yields the state associated to
the "vacuum" of the
conformal field theory $G_k$. Explicitly,
$$\Gamma = \Sigma_{w\in W}\ (-1)^w \Pi_{j}
B_j^{{w(\rho)}_j}      \ . \eqno(B.4)$$
where $W$ is the Weyl group, $B_j$ are, as before, the raising operator
associated with the co-root to $\alpha_j$ and $\rho$ is
one-half the sum of the positive weights.
Note that the exponent ${w(\rho)}_j$ is the $j$-th component of the
Weyl transform of $\rho$ but written in the co-root basis (for
further details consult Refs.[18,19].)

Trivially, each  $\Gamma$ (and therefore $K^{-1}$) commutes with
any polynomial in the $B_j$'s. What we need to show is that
$\Gamma^\dagger\Gamma$ commutes with all $z_1=\Pi_{j} A_j^{s_j}$
for the $s_j$ given in Eq.(B.3). Then the full group of Wilson
lines on the torus associated to the center action of $G$ will
commute with $K^{-1}$. In view of the commutation relations of the
Gaussian model, application of Schur's lemma yields
$$ \big[z_1,\Pi_{j} B_j^{{w(\rho)}_j}\big]=exp\big({{2\pi i}\over{k+c}}
s_l (C^{-1})_{lm}w(\rho)_m\big)\ \ , \eqno(B.5)$$
where $w(\rho)_m$ are the components of $w(\rho)$ in the co-root basis.
Note that for any simple root $\alpha$, $w_\alpha(\rho)=\rho-\alpha$.
Now since we go from the root basis to the co-root basis by the
application of the matrix $C$, then by Eq.(B.3) we see that
the right hand side of Eq.(B.5) is independent of which
Weyl transformation $w$ appears on the left hand side.
Thus,
$$ \big[z_1,\Gamma]=exp\big({{2\pi i}\over{k+c}}s_l (C^{-1})_{lm}\rho_m\big)
\ \ .\eqno(B.6)$$
Finally, since $\rho={{1}\over{2}}\Sigma_{\alpha>0}\alpha$ then
as a vector in the root basis $\rho$ has components that are
integer and half-integer. Thus, the right hand side of Eq.(B.6)
is either $+1$ or $-1$, depending essentially on the group and the
particular $z_1$. Therefore
$$\big[z_1,\Gamma^\dagger\Gamma\big]=1 \ \ . \eqno(B.7)$$
and so all Wilson lines on the torus
that represent the center action in $G_k$
commute with $K^{-1}$.
\end